\documentclass[a4paper]{jpconf}
\usepackage{graphicx}
\usepackage{wallpaper}

\begin{document}
\title{Electron localization properties in high pressure hydrogen at the liquid-liquid phase transition by
Coupled Electron-Ion Monte Carlo}

\author{Carlo Pierleoni}
\address{Department of Physical and Chemical Sciences, University of L'Aquila, Italy\\
Maison de la Simulation, CEA, CNRS, Univ. Paris-Sud, UVSQ, Universit\'e Paris-Saclay, 91191 Gif-sur-Yvette, France}

\author{Giovanni Rillo}
\address{Department of Physics, Sapienza University of Rome, Italy}

\author{David M.Ceperley}
\address{Department of Physics, University of Illinois Urbana-Champaign, Illinois, USA}

\author{Markus Holzmann}
\address{LPMMC, UMR 5493 of CNRS, Universit\'e Grenoble Alpes, France\\
Institut Laue-Langevin, BP 156, F-38042 Grenoble Cedex 9, France}

\begin{abstract}
We analyze in detail the electronic properties of high pressure hydrogen around the 
liquid-liquid phase transition based on Coupled Electron-Ion Monte Carlo calculations.
Computing the off-diagonal single particle density matrix and the momentum distribution
we discuss localization properties of the electrons. The abrupt changes of these distributions
indicate a metal to insulator transition occurring together with the structural transition
from the atomic to molecular fluid. We further discuss the electron-proton and
electron-electron pair correlation functions, which also change abruptly at the transition.
\end{abstract}

\section{Introduction}
The search of metallization in hydrogen has been and still is one of the major 
driving forces in high pressure research. In 1935,  Wigner and Huntington discussed
the processes of metallization of low temperature solid molecular hydrogen upon
increasing pressure either by a continuous process or through a structural phase transition
to a monoatomic phase. They predicted that a body centered cubic monoatomic hydrogen would
become favorable over the molecular form above 25GPa.
Despite almost a century of intense research and impressive progress in both experimental
 and theoretical methods, this question still remains open.
Static pressure experiments by the Diamond Anvil Cell technique in solid hydrogen have reported
the record pressure of $\sim$500GPa \cite{Dias2017,Eremets2017} and the emergence of metallic
 or semi-metallic behaviors, but the mechanism of metallization is still unclear.
There has been intense interest in this system during the past year with the first
claim of experimental observation of metallic solid hydrogen, however followed by
many criticisms \cite{Dias2017,Goncharov2017,Loubeyre2017,Liu2017,Silvera2017}.
On the theoretical side,
{\it ab initio} DFT calculations using the most common exchange-correlation approximations,
have not been found to be accurate enough to address the ordering of the phases of hydrogen near metallization
and molecular dissociation \cite{Clay2014,Azadi2013b}. Only recently, more fundamental methods
based on Quantum Monte Carlo approaches, together with adequate computer resources,
have become available and have started to be used to investigate potential phases of
crystalline hydrogen \cite{Drummond2015,Rillo2017}. We expect that in the near future those methods
will contribute to clarify the scenario of solid hydrogen metallization 
and provide fundamental information for interpreting the experimental results. 

The situation seems slightly easier in the higher temperature fluid, at least concerning
theoretical methods. Experimentally, a discontinuous change in the reflectivity and the resistivity
of fluid hydrogen has been observed, both with shock wave and static compression
techniques  \cite{Weir1996,Knudson2015,Dzyabura2013,Zaghoo2016}. However, the location of this
discontinuity and its nature is still under debate, results from these two different methods
differing by as much $\sim$150GPa. Theoretical {\it ab initio} calculations also predict a discontinuity
in the Equation of State, a first order phase transition between an insulating molecular fluid
and a dissociated metallic fluid. Again, different approximations predict different locations
of the transition line \cite{Pierleoni2016} with the most accurate predictions based on
Quantum Monte Carlo (QMC)\cite{Pierleoni2016, Pierleoni2017, Mazzola2017} being rather close
to the results from the static compression method,
although at slightly higher pressures \cite{Pierleoni2016}.

These simulations also allow the characterization of the nature of the phase transition as
concerns both nuclear and electronic properties. We have shown that at the transition,
molecules suddenly dissociate and the electronic localization changes, accompanied by the
emergence of a non-zero static electronic conductivity (the latter obtained at the DFT level) \cite{Pierleoni2016}. While we have recently extended and integrated the description of the change of nuclear
structure at the transition\cite{Pierleoni2017}, we have only given a rapid account of the
electronic properties at the transition in the original publication \cite{Pierleoni2016}. In the
present paper we provide a more complete account of the changes in the electronic properties
leading to a coherent theoretical picture of the liquid-liquid phase transition in hydrogen.

The paper is organized as follows. In the next section, after a brief recap of the basic
ingredients of the Coupled Electron-Ion method, we present a detailed account of the theory and practical implementation of single electron density matrix calculations
and the electronic momentum distribution used to characterize the electronic localization
properties. We then discuss the results of ref. \cite{Pierleoni2016} in more 
detail and  show original results for the density-density, electron-proton
and electron-electron, pair distribution functions. 

\section{Method}
\subsection{The Coupled-Electron-Ion Monte Carlo Method}
The Coupled Electron-Ion Monte Carlo method is an {\it ab initio} method for finite temperature nuclei coupled with ground state electrons based on the Born-Oppenheimer approximation.
It is described in detail in refs. \cite{Pierleoni2006,McMahon2012a}. More recent implementations
and technical details employed in the present calculations are provided in the Supplementary 
Information of ref. \cite{Pierleoni2016} together with a detailed discussion of the
various systematic biases introduced by numerical approximations. 
In the following, we only discuss the methodological issues concerning the calculation
of the reduced single particle density matrix and momentum distribution which have never
been addressed in detail before, in particular 
the connection with twisted boundary conditions to reduce finite size effects.

\subsection{The reduced single particle density matrix and momentum distribution}

For any given many-body state of the electrons,
$\Psi({\bf R}_N)$, ${\bf R}\equiv ({\bf r}_1,{\bf r_2}, \dots, {\bf r}_N)$,
all observables involving only one particle can be calculated from the knowledge of
the reduced single particle density matrix 
\begin{equation}
\rho^{(1)}({\bf r},{\bf r}') = N \frac{\int d {\bf r}_2 \cdots d{\bf r}_N
\Psi^*({\bf r},{\bf r}_2,\dots,{\bf r}_N)
\Psi({\bf r}',{\bf r}_2,\dots,{\bf r}_N)}{\int d{\bf R}_N |\Psi({\bf R}_N) |^2} 
\end{equation}
The real space density distribution of the electrons is given by its diagonal elements, $\rho({\bf r})\equiv \rho^{(1)}({\bf r},{\bf r})$. 
From a Monte Carlo sampling of the probability $\Pi({\bf R}_N) \propto |\Psi({\bf R}_N)|^2$  we can calculate
the off-diagonal elements via reweighting
\begin{equation}
\rho^{(1)}({\bf r},{\bf r}') = N \left\langle \delta({\bf r} -{\bf r}_1) 
\frac{\Psi({\bf r}',{\bf r}_2,\dots,{\bf r}_N)}
{\Psi({\bf r},{\bf r}_2,\dots,{\bf r}_N)} \right\rangle
\end{equation}
where
$\langle \cdot \rangle
 \equiv \int d {\bf R}_N   \Pi({\bf R}_N) [\cdot ]$. Whereas the
reduced single particle density matrix is rather directly calculated within
variational Monte Carlo \cite{McMillan1965}, the sampling must be adapted to obtain
unbiased estimators within reptation Monte Carlo \cite{Moroni2004,Holzmann2011}.

\begin{figure}[h]
\begin{center}
\includegraphics[width=\columnwidth]{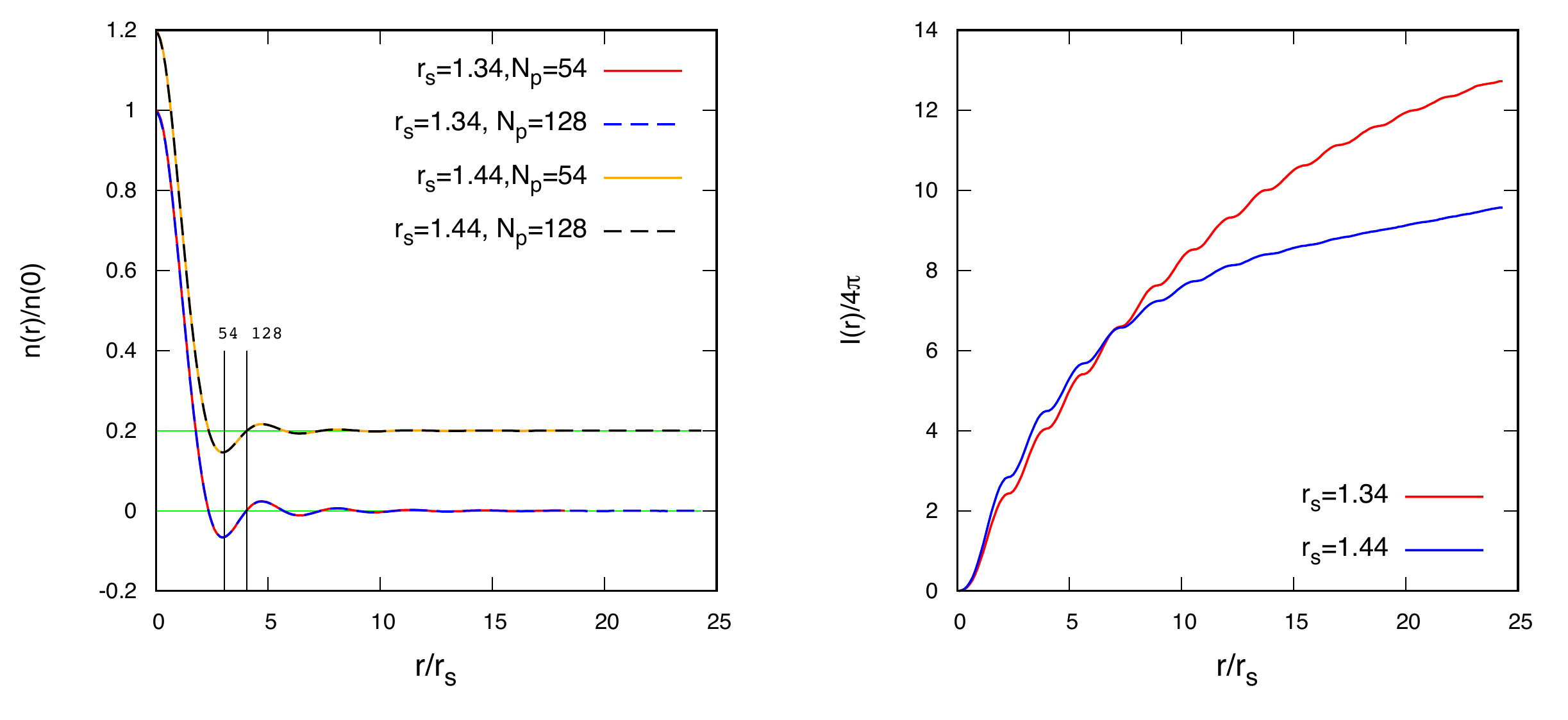}
\end{center}
%\hspace{2pc}%
\caption{\label{nr}Left panel: The off-diagonal single particle density matrix, $n(r)$, at $T=1200$K and two different densities $r_s=1.34$ in the atomic and $r_s=1.44$ in the molecular phase (with an offset of $0.2$) of the liquid-liquid phase transition. 
Vertical lines indicate $L/2$, where $L$ is the length of the simulation cell, for systems containing $N_p=54$ and $N_p=128$ protons.
The use of twisted boundary conditions gives access to $n(r)$ at distances $r >> L/2$, without noticeable bias. 
Right panel: Convergence of the integral of the absolute value of the off-diagonal single particle density matrix, $I(r)=4 \pi \int_0^{r}dr' r'^2 |n(r')|/n(0)$, with respect to the distance $r$ at $T=1200$K.
Its behavior at large distance provides a measure of electron localization in real space. Divergence occurs for delocalized, extended wave functions, typical for metallic systems, whereas the integral should asymptotically converge when electrons are localized (insulators).
Although the region of convergence of the integral of the molecular fluid is not fully reached
(despite the extended range accessed by the use of twisted boundary conditions), the atomic and molecular fluid are clearly distinguished.
}
\end{figure}

Diagonalizing the reduced single particle density matrix we obtain the natural orbitals. 
For homogeneous systems, these are plane waves and $\rho^{(1)}({\bf r},{\bf r}')$
depends only on the distance $|{\bf r}-{\bf r}'|$. Since averaging over the proton configuration
restores translational invariance in the fluid phase of hydrogen, we introduce the volume--averaged off-diagonal density matrix
\begin{equation}
n(r)=\frac{1}{V} \int d{\bf r}' \rho^{(1)}({\bf r}+{\bf r}',{\bf r}')
\end{equation}
whose Fourier transform gives the momentum distribution
\begin{equation}
n_k= \int d {\bf r} e^{-i {\bf k} \cdot {\bf r}} n(r)
\end{equation}
These are normalized such that $n(0)=V^{-1}\sum_{\bf k} n_k$ gives the average electron
density $\rho=3/(4 \pi r_s^3) a_B^{-3}$, usually given in terms of the dimensionless parameter
$r_s$, where $a_B$ is the Bohr radius.

For a normal Fermi liquid such as the homogeneous disorder-free electron gas,
the momentum distribution 
is characterized by a discontinuity of the momentum distribution 
at the Fermi wave vector \cite{Huotari2010,Holzmann2011}, leading to oscillations in $n(r)$ whose envelop decays algebraically as $r^{-3}$ at large distances.

\subsection{Twist averaged boundary conditions}

Using twisted boundary conditions for the wave functions \cite{Lin2001}, we have
\begin{equation}
\Psi_{\theta}({\bf r}_1 + L \hat{{\bf x}}_\alpha,{\bf r}_2,\dots)=e^{i \theta_\alpha}
\Psi_{\theta}({\bf r}_1,{\bf r}_2,\dots)
\end{equation}
where ${\bf x}_\alpha$ is the unit vector in the $\alpha$ direction ($\alpha=x,y,z$), and
$\theta_\alpha$ is the twist angle in this direction,
$-\pi \le \theta_\alpha < \pi$.
For a general twist angle, we then have for the reduced density matrix
\begin{equation}
\rho^{(1)}_\theta({\bf r}+L {\bf x}_\alpha,{\bf r}')
= e^{i \theta_\alpha} \rho^{(1)}_\theta({\bf r},{\bf r}')
\end{equation}
Averaging over $N_\theta$ different twist angles, 
we then get for the twist-averaged off-diagonal density matrix,
\begin{equation}
n_{\mathrm{TABC}}({\bf r}+ L \hat{{\bf x}}_\alpha)= \frac{1}{N_\theta} \sum_{i=1}^{N_\theta} e^{i \theta_\alpha} n_{\phi_i}({\bf r})
\end{equation}
Whereas $n_{\phi_i}({\bf r})$ is only defined within the simulation box,
the twist averaged density matrix can thus be extended to distances much larger than $L$ by
using a fine grid for the twist angles. 

In Fig.~\ref{nr}, we show the twist averaged density matrix which smoothly extends the behavior of the off-diagonal matrix elements to distances much larger than $L/2$, 
the typical limit for simulations using strictly periodic boundary conditions.
Comparison of two different system sizes does not show any noticeable bias. However,
we note that slowly decaying finite-size effects due to two-particle correlations as described in Refs \cite{Holzmann2009,Holzmann2011} are not taken into account by twist averaged boundary conditions and have
to be addressed separately.

\begin{figure}
\begin{center}
\includegraphics[width=1.05\columnwidth]{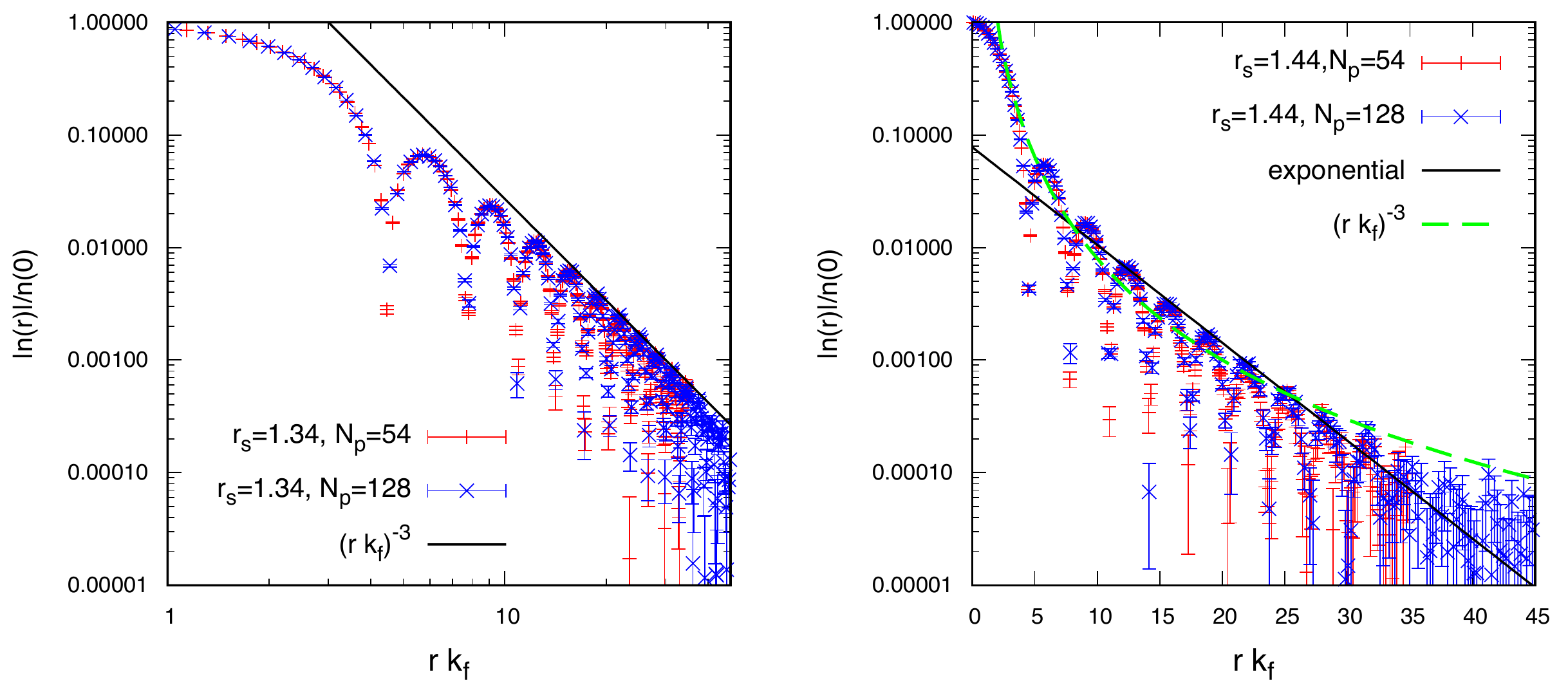}
\end{center}
\caption{\label{nrlog}Absolute value of the off-diagonal part of the reduced single particle density matrix at $T=1200$K in a logarithmic scale. For the atomic liquid at $r_s=1.34$ (left panel), the decay of the
envelope asymptotically decays as $(k_Fr)^{-3}$, whereas it is closer to an exponential behavior for the molecular liquid at $r_s=1.44$ (right panel). }
\end{figure}

Using a linear grid of twist angles, $\theta_\alpha=2 \pi n/ML$ with integer $n$, 
the twist averaged density matrix,
$n_{\mathrm{TABC}}({\bf r})$, is a periodic function in a box of extension $ML$ with a discrete
Fourier transform on the reciprocal lattice ${\bf k}=(n_x,n_y,n_z) 2 \pi/ML$.
The corresponding momentum distribution writes
\begin{equation}
n_{\bf k}^{\mathrm{TABC}}= \int_{M^3 V} d {\bf r} e^{- i {\bf k} \cdot {\bf r}} n_{\mathrm{TABC}}({\bf r})
= \frac{1}{N_\theta} \sum_{i=1}^{N_\theta} \sum_{\bf m} \int_V d{\bf r} e^{i\sum_\alpha\left[ \theta_\alpha m_{\alpha} -  k_\alpha ({\bf r}_{\alpha}+m_{\alpha}L)\right] }
n_{\phi_i}({\bf r})
\end{equation}
The summation over the integer values $m_\alpha=0,\dots M-1$ gives a periodic delta function which
selects a unique twist angle $\theta_\alpha$ modulus $2\pi$.
We can therefore write ${\bf k}={\bf k}_0 +{\bf G}$ where ${\bf G}=(n_x,n_y,n_z) 2 \pi/L$ is on the reciprocal lattice 
of the original system of extension $L$, and write
\begin{equation}
n_{\bf k}^{\mathrm{TABC}}=\int_V d{\bf r} e^{-i({\bf k}_0+{\bf G}) \cdot {\bf r}} n_{\theta}({\bf r})
\end{equation}
Therefore, the natural extension of the momentum distribution from periodic to twisted boundary conditions is given by
\begin{equation}
n_{{\bf G}+{\bf k}_0}=\int_V d{\bf r} e^{-i({\bf k}_0+{\bf G}) \cdot {\bf r}} n_{\theta}({\bf r})
\end{equation}
where $k_0=\theta/L$ shifts the grid in reciprocal space with respect to simulations
using periodic boundary conditions. Twist averaged boundary conditions therefore allow us to 
extend the momentum distribution to an arbitrarily dense grid in reciprocal space.
To impose a sharp Fermi surface, simulations should be done at fixed chemical potential using
grand-canonical twist averaging 
where the number of electrons may vary with
twist angle  \cite{Chiesa2006,Holzmann2016}.

\section{Results}

In the following we analyze the electronic properties of Coupled-Electron Ion Monte Carlo calculations \cite{Pierleoni2016} along five isotherms $T=600K, 900K, 1200K, 1500K, 3000K$.
For all temperatures $T \le 1500K$, a weak first-order transition from a molecular to an atomic liquid, together with an insulator to metal transition, occurs upon increasing density \cite{Pierleoni2016}.
In a recent paper, we have analyzed in details the structural properties of the protons, 
confirming the abrupt character of the molecular dissociation at the transition by employing a cluster analysis of the nuclear configurations visited during the CEIMC runs \cite{Pierleoni2017}.
In the following, we analyze the electronic properties averaged over the nuclear configurations in more detail, and show that the change in the structural properties is accompanied with a sharp 
change of the electronic single particle density matrix as well as the electron-proton and electron-electron pair correlation functions.
All the following calculations of the off-diagonal density matrix
are done at the VMC level using twisted boundary conditions with fixed number of electrons
(canonical twist averaging). These results may still suffer from size effects 
due to two-particle correlations \cite{Holzmann2009,Holzmann2011} which decays slowly
with increasing system size. However, these effects will not affect our main result showing
a clear qualitative change in the off-diagonal single particle density matrix
from the molecular to the atomic fluid.

\begin{figure}
\begin{center}
\includegraphics[width=\columnwidth]{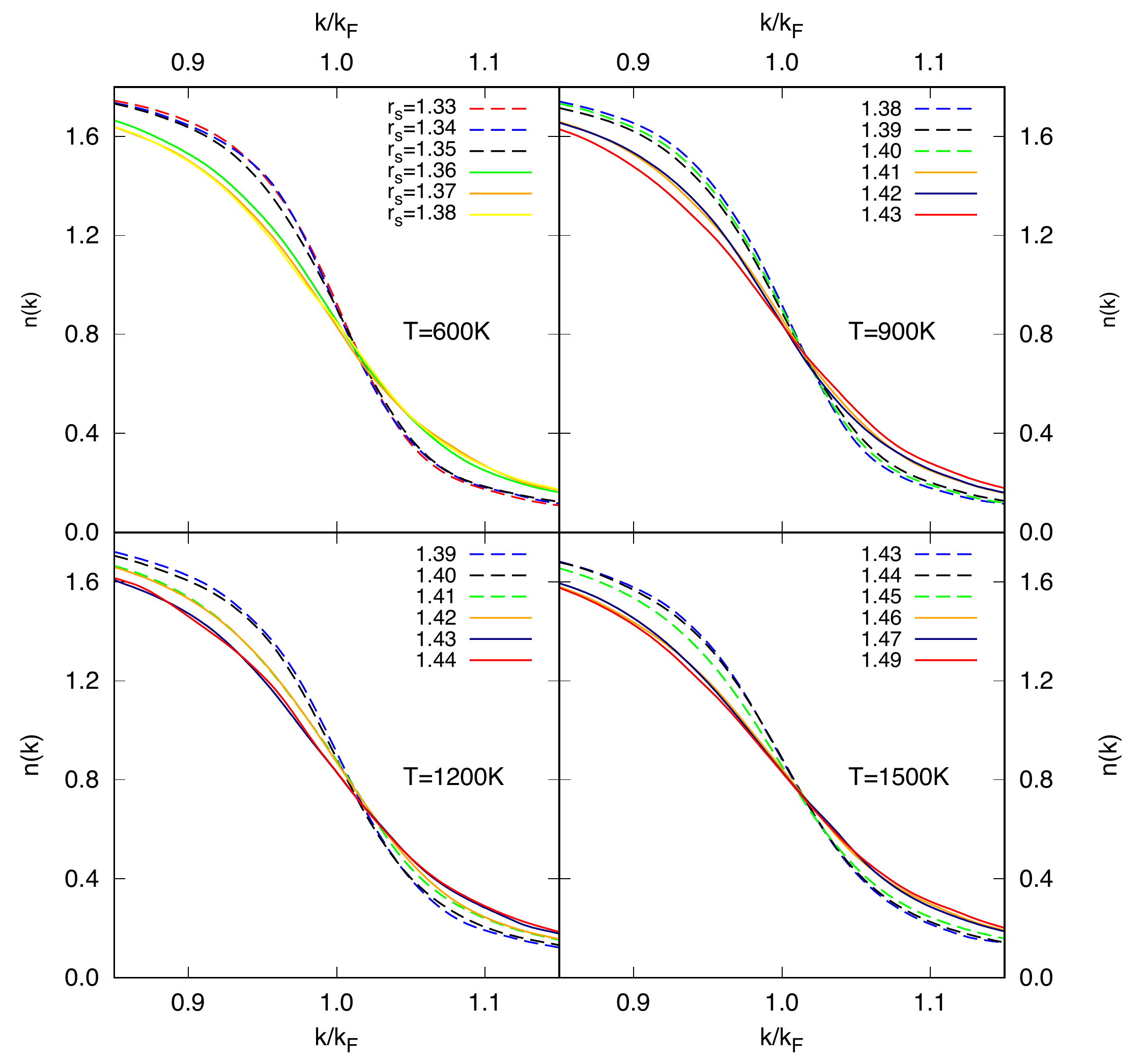}
\end{center}
\caption{\label{nk}Momentum distribution, $n(k)$, of the electrons at different temperatures 
and densities. The distributions of the atomic fluid (dashed lines) are clearly distinguished from those of the molecular ones (full lines), except at the coexisting densities.}
\end{figure}

\subsection{The off-diagonal density matrix, electron localization, and momentum distribution}

In Figure \ref{nr} and \ref{nrlog}, we show the off-diagonal elements of the reduced single particle density matrix, $n(r)$, for the atomic ($r_s=1.34$) and molecular ($r_s=1.44$) liquid on both
sides of the phase transition at $T=1200K$. At first glance the two behaviors are qualitatively similar, but
because of the logarithmic scale used in Fig. \ref{nrlog}, we see that the amplitude
of the oscillations decays roughly exponentially in the molecular phase, whereas it is much closer
to a Fermi-liquid like algebraic decay $\sim r^{-3}$ in the atomic phase. 

\begin{figure}
\begin{center}
\includegraphics[width=\columnwidth]{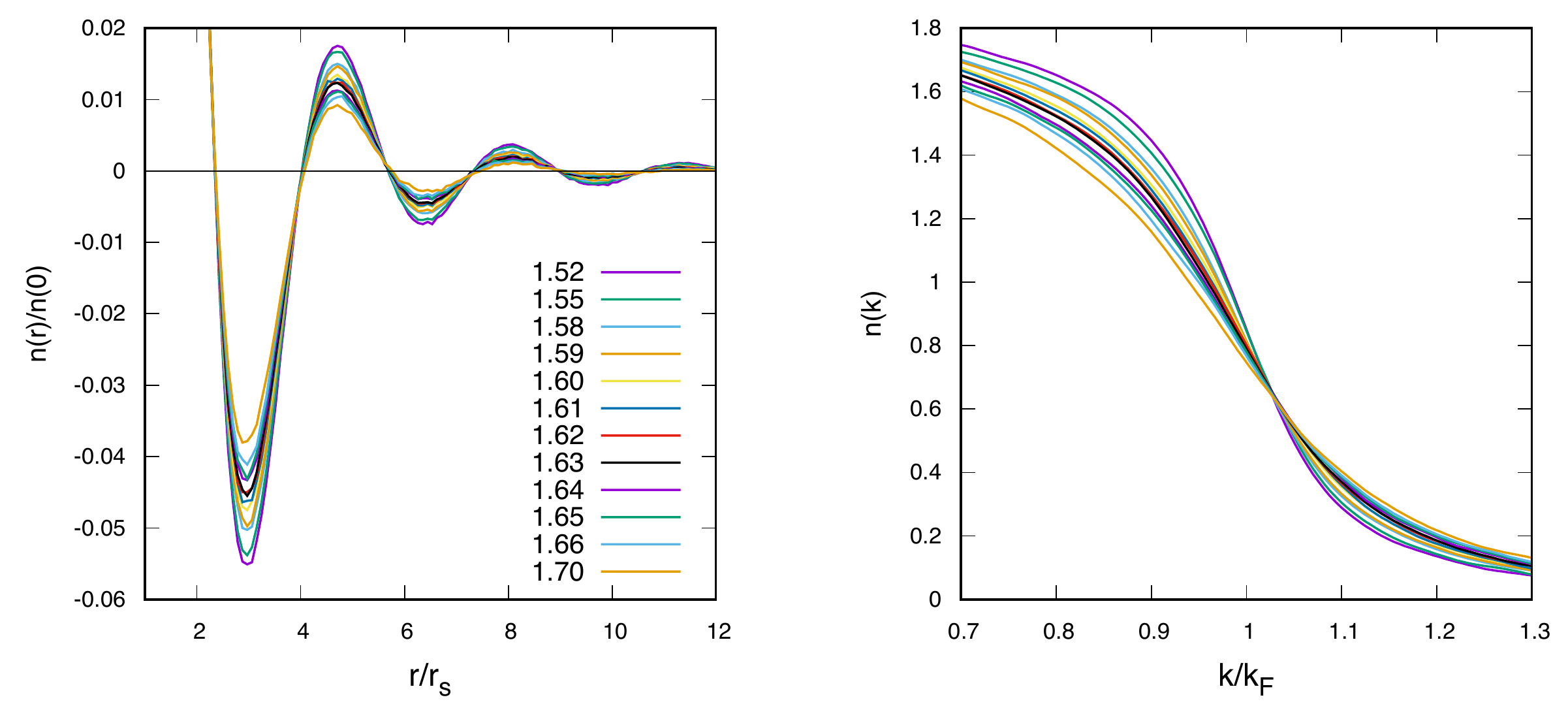}
\end{center}
\caption{\label{t3000}Off-diagonal density matrix, $n(r)$, and momentum distribution, $n(k)$, of the electrons at $T=3000$K, above the critical point of the liquid-liquid phase transition.
Both distributions change smoothly with density without indication of a localization transition.
}
\end{figure}

Integration of the absolute value of $n({\bf r})$ introduces a measure of electron localization \cite{Mazzola2015,Pierleoni2016}. For any decay faster than $r^{-3}$, the integral 
$I(r)=4 \pi \int_0^{r}dr' r'^2 |n(r')|/n(0)$ converges with respect to $r \to \infty$ and
single particle density matrix is expected to be localized in space
leading to insulating character, otherwise metallic properties should occur due to the extended character
of the single particle density matrix. As we can see from Fig.~\ref{nr}, the integral indeed seems to diverge on the atomic side of the transition, whereas it may converge for the molecular fluid.
However, despite the extended region reached by the use of twisted boundary conditions, the integral
has not yet obtained its asymptotic value.

The drastic change of the electronic single particle properties at the phase transition is even
more evident in the momentum distribution, $n(k)$, shown in Fig.~\ref{nk} for the four 
isotherms below the critical temperature. 
Along each isotherm, we report results for 6 densities (expressed in terms of $r_s$), around the transition point as done in ref \cite{Pierleoni2017}.
In terms of $k/k_F$, where $k_F=(9\pi/4)^{1/3} (r_s a_B)^{-1}$ is the Fermi vector corresponding to the
average electronic density, the distribution in Fourier space changes very smoothly with density or temperature 
within a single phase. However, the distribution changes rapidly at the transition, that of atomic liquid
is substantially different from that of the molecular one\footnote{At T=1200K $n(k)$'s for the two coexisting densities are almost superimposed. Approaching the critical temperature the system spontaneously switch from one phase to the other along the trajectory resulting in an average behavior for long sampling. We believe this is the reason for this superimposition. At T=1500K, the generated trajectories are rather shorter.}.

\begin{figure}
\begin{center}
\includegraphics[width=\columnwidth]{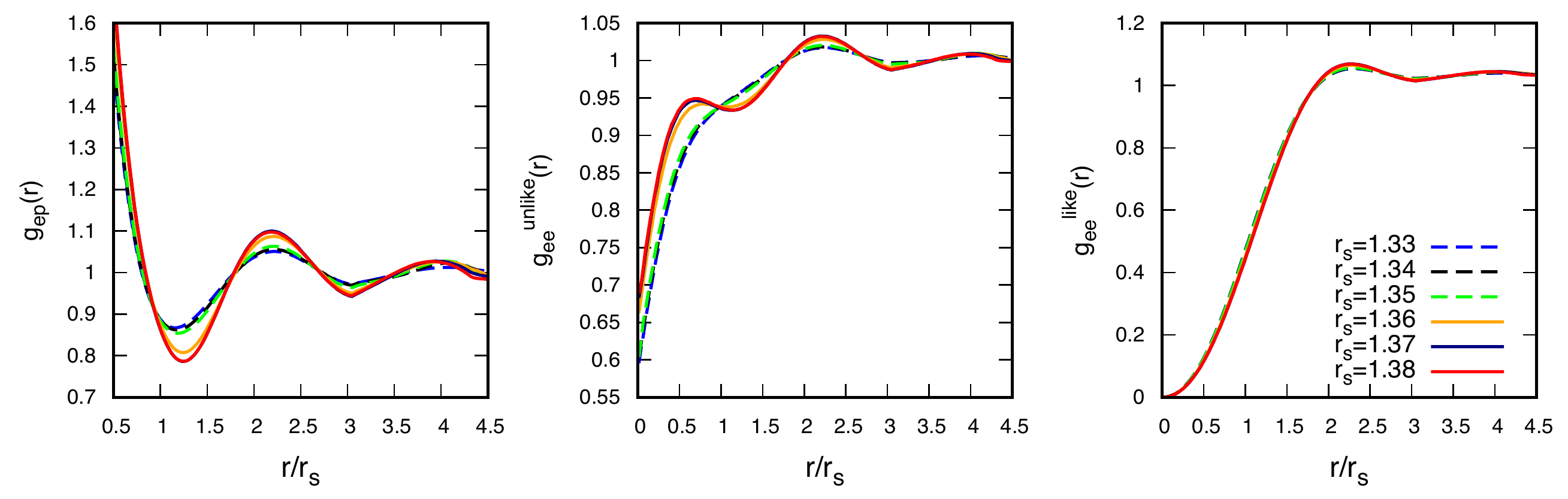}
\end{center}
\caption{\label{gr600}Electron-proton and electron-electron pair correlation with unlike and like spin, 
$g_{ep}(r)$, $g_{ee}^\mathrm{unlike}(r)$, and $g_{ee}^\mathrm{like}(r)$, respectively, at $T=600K$.
The distributions abruptly change at the transition from the molecular to the atomic fluid.}
\end{figure}

Although the momentum distribution at $k_F$ is much steeper for the atomic fluid, there
is not a discontinuity at the Fermi surface that should be expected for a normal Fermi fluid.
However, in our case, even in the metallic state, the electrons collide with protons, which
can be considered as a static external potential within the Born-Oppenheimer approximation.
In contrast to electrons moving in a periodic potential, even in the non-interacting
electron approximation, the mean-free path  (and life-time) 
of the electrons in the disorder potential created
by the protons becomes finite and the discontinuity of the momentum distribution at the Fermi 
surface gets smoothed out.

At a higher temperature, $T=3000K$, shown in Fig.~\ref{t3000}, the off-diagonal density matrix as well as the momentum distribution changes continuously with density without any sign of a phase transition in the electronic properties, and
we do not expect the occurrence of the metal to insulator transition at this temperature
in agreement with the absence of a phase transition in the
the structural analysis of the nuclear positions and in the equation of state \cite{Pierleoni2016,Pierleoni2017}.
Therefore, the transition in the electronic properties, from a localized (insulating) to an
extended (metallic) state coincides,
within our precision, with the structural transition
from the molecular to the atomic liquid at all temperatures.

\subsection{The density-density correlation functions}

\begin{figure}
\begin{center}
\includegraphics[width=\columnwidth]{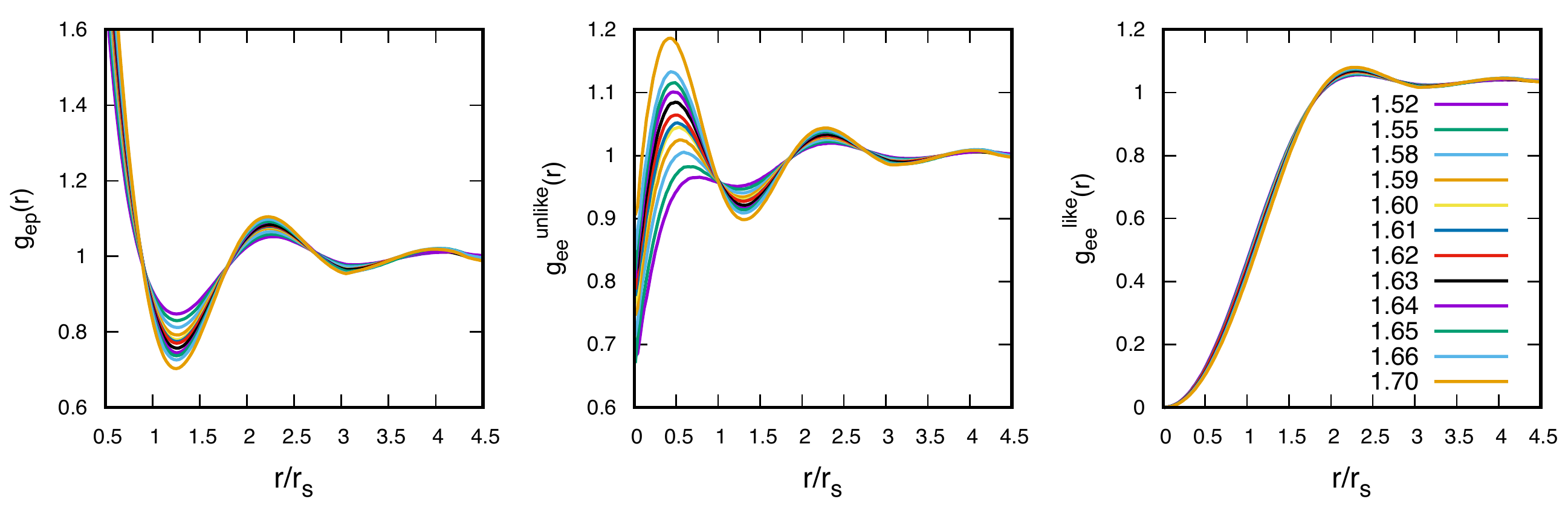}
\end{center}
\caption{\label{gr3000}Electron-proton and spin-resolved electron-electron correlation functions,
$g_{ep}(r)$, $g_{ee}^\mathrm{unlike}(r)$, and $g_{ee}^\mathrm{like}(r)$, respectively, at $T=3000K$.
Here, above the critical temperature of the structural transition, the distribution changes smoothly with density.}
\end{figure}

The electron-proton and electron-electron correlation functions at low temperature are shown in Fig.\ref{gr600}. Similar to the off-diagonal single particle
properties discussed above, there is an abrupt change of the distributions at the liquid-liquid
phase transition, more pronounced for unlike than for like spins in the electron-electron 
correlations. However, whereas the off-diagonal density matrix
properties enter into transition matrix elements, e.g. needed for transport calculations, 
density-density correlation functions reflect structural properties.
The changes in the electron-proton and unlike electron-electron
correlation functions at the transition therefore have a rather direct interpretation in terms
of the change from atomic to molecular bonds.
Above the critical temperature, the pair correlation functions change smoothly with density
(see Fig.\ref{gr3000}) again signalling the absence of a first order phase transition.

\section{Conclusions}

In this paper, we have presented the methodology for calculating off-diagonal density matrices using twisted boundary conditions and analyzed the changes in the electronic properties of hydrogen 
at the liquid-liquid phase transition. Our analysis shows an abrupt change in all electronic
properties, indicating an insulator to metal transition 
which occurs simultaneously with the structural transition from the molecular to atomic
fluid.
At higher temperatures, $T=3000K$, above the critical point of the first-order transition, 
electronic properties change smoothly with density.

\subsection{Acknowledgments}

C.P. was supported by the Agence Nationale de la Recherche (ANR) France, under the program ``Accueil de Chercheurs de Haut Niveau 2015'' project: HyLightExtreme. D.M.C. was supported by DOE Grant NA DE-NA0001789 and by the Fondation NanoSciences (Grenoble).  Computer time was provided by PRACE Projects 2013091918 and 2016143296 and by an allocation of the Blue Waters sustained petascale computing project, supported by the National Science Foundation (Award OCI 07- 25070) and the State of Illinois.  
\section*{References}
\bibliography{dottorato}

\providecommand{\newblock}{}
\begin{thebibliography}{10}
\expandafter\ifx\csname url\endcsname\relax
  \def\url#1{{\tt #1}}\fi
\expandafter\ifx\csname urlprefix\endcsname\relax\def\urlprefix{URL }\fi
\providecommand{\eprint}[2][]{\url{#2}}

\bibitem{Dias2017}
Dias R~P and Silvera I~F 2017 {\em Science\/} {\bf 355} 715--718

\bibitem{Eremets2017}
Eremets M, Drozdov A~P, Kong P and Wang H 2017  (\textit{Preprint}
  \eprint{https://arxiv.org/abs/1708.05217})
  \urlprefix\url{https://arxiv.org/abs/1708.05217}

\bibitem{Goncharov2017}
Goncharov A~F and Struzhkin V~V 2017  (\textit{Preprint}
  \eprint{arXiv:1702.04246})
  \urlprefix\url{https://arxiv.org/pdf/1702.04246.pdf}

\bibitem{Loubeyre2017}
Loubeyre P, Occelli F and Dumas P 2017  (\textit{Preprint}
  \eprint{arXiv:1702.07192})
  \urlprefix\url{https://arxiv.org/pdf/1702.07192.pdf}

\bibitem{Liu2017}
Liu X~D, Dalladay-Simpson P, Howie R~T, Li B and Gregoryanz E 2017
  (\textit{Preprint} \eprint{arXiv:1704.07601v2})
  \urlprefix\url{https://arxiv.org/pdf/1704.07601.pdf}

\bibitem{Silvera2017}
Silvera I~F and Dias R 2017  (\textit{Preprint} \eprint{arXiv:1703.03064})
  \urlprefix\url{https://arxiv.org/pdf/1703.03064.pdf}

\bibitem{Clay2014}
Clay R~C, McMinis J, McMahon J~M, Pierleoni C, Ceperley D~M and Morales M~A
  2014 {\em Phys. Rev. B\/} {\bf 89} 184106

\bibitem{Azadi2013b}
Azadi S and Foulkes W~M~C 2013 {\em Phys Rev. B\/} {\bf 88} 014115

\bibitem{Drummond2015}
Drummond N~D, Monserrat B, Lloyd-Williams J~H, R{\'{i}}os P~L, Pickard C~J and
  Needs R~J 2015 {\em Nature Communications\/} {\bf 6} 7794

\bibitem{Rillo2017}
Rillo G, Morales M~A, Ceperley D~M and Pierleoni C 2018 {\em J. Chem. Phys.\/}
  {\bf 148} 102314

\bibitem{Weir1996}
Weir S~T, Mitchell A~C and Nellis W~J 1996 {\em Phys. Rev. Letts.\/} {\bf 76}
  1860--1863 ISSN 0031-9007
  \urlprefix\url{http://link.aps.org/doi/10.1103/PhysRevLett.76.1860}

\bibitem{Knudson2015}
Knudson M~D, Desjarlais M~P, Becker A, Lemke R~W, Cochrane K~R, Savage M~E,
  Bliss D~E, Mattsson T~R and Redmer R 2015 {\em Science\/} {\bf 348}
  1455--1460 ISSN 0036-8075
  \urlprefix\url{http://www.sciencemag.org/cgi/doi/10.1126/science.aaa7471}

\bibitem{Dzyabura2013}
Dzyabura V, Zaghoo M and Silvera I~F 2013 {\em Proc. Nat. Acad. Sc.\/} {\bf
  110} 8040--8044

\bibitem{Zaghoo2016}
Zaghoo M, Salamat A and Silvera I~F 2016 {\em Phys. Rev. B\/} {\bf 93} 155128
  ISSN 1550235X

\bibitem{Pierleoni2016}
Pierleoni C, Morales M~A, Rillo G, Holzmann M and Ceperley D~M 2016 {\em Proc.
  Natl. Acad. Sci.\/} {\bf 113} 4954--4957 ISSN 0027-8424
  \urlprefix\url{http://www.pnas.org/lookup/doi/10.1073/pnas.1603853113}

\bibitem{Pierleoni2017}
Pierleoni C, Holzmann M and Ceperley D 2018 {\em Contribution to Plasma
  Physics\/}  accepted

\bibitem{Mazzola2017}
Mazzola G, Helled R and Sorella S 2017  (\textit{Preprint}
  \eprint{arXiv:1709.08648})

\bibitem{Pierleoni2006}
Pierleoni C and Ceperley D 2006 {The coupled electron-ion monte carlo method}
  {\em Lecture Notes in Physics\/} vol 703 (Springer, Berlin) pp 641--683 ISBN
  3540352708 | 9783540352709

\bibitem{McMahon2012a}
McMahon J~M, Morales M~A, Pierleoni C and Ceperley D~M 2012 {\em Rev. Mod.
  Phys.\/} {\bf 84} 1607--1653 ISSN 0034-6861
  \urlprefix\url{http://link.aps.org/doi/10.1103/RevModPhys.84.1607}

\bibitem{McMillan1965}
McMillan W 1965 {\em Phys. Rev.\/} {\bf 138} 442--451 ISSN 0031-899X
  \urlprefix\url{http://journals.aps.org/pr/abstract/10.1103/PhysRev.138.A442}

\bibitem{Moroni2004}
Moroni S and Boninsegni M 2004 {\em J. Low Temp. Phys.\/} {\bf 136} 129

\bibitem{Holzmann2011}
Holzmann M, Bernu B, Pierleoni C, McMinis J, Ceperley D~M, Olevano V and {Delle
  Site} L 2011 {\em Phys. Rev. Letts.\/} {\bf 107} 1--5 ISSN 00319007
  (\textit{Preprint} \eprint{1105.2338})

\bibitem{Huotari2010}
Huotari S, Soininen J~A, Pylkk{\"a}nen T, H{\"a}m{\"a}l{\"a}inen K, Issolah A,
  Titov A, McMinis J, Kim J, Esler K, Ceperley D~M, Holzmann M and Olevano V
  2010 {\em Phys. Rev. Lett.\/} {\bf 105} 086403

\bibitem{Lin2001}
Lin C, Zong F~H and Ceperley D~M 2001 {\em Physical review. E\/} {\bf 64}
  016702 ISSN 15393755 (\textit{Preprint} \eprint{0101339})

\bibitem{Holzmann2009}
Holzmann M, Bernu B, Olevano V, Martin R~M and Ceperley D~M 2009 {\em Phys.
  Rev. B\/} {\bf 79} 2--5 ISSN 10980121 (\textit{Preprint} \eprint{0810.2450})

\bibitem{Chiesa2006}
Chiesa S, Ceperley D~M, Martin R~M and Holzmann M 2006 {\em Phys. Rev.
  Letts.\/} {\bf 97} 6--9 ISSN 00319007

\bibitem{Holzmann2016}
Holzmann M, Clay R, Morales M~A, Tubmann N~M, Ceperley D~M and Pierleoni C 2016
  {\em Phys. Rev. B\/} {\bf 94} 035126 ISSN 2469-9950
  \urlprefix\url{http://link.aps.org/doi/10.1103/PhysRevB.94.035126}

\bibitem{Mazzola2015}
Mazzola G and Sorella S 2015 {\em Phys. Rev. Letts.\/} {\bf 114} 1--5

\end{thebibliography}
%\begin{thebibliography}{9}

%\bibitem{McMillan} W. L. McMillan, Phys. Rev. {\bf 138}, A442 (1965).
%\bibitem{Saverio} S. Moroni and M. Boninsegni, J. Low Temp. Phys. {\bf 136}, 129 (2004).
%\bibitem{momk3D} M. Holzmann, B. Bernu, C. Pierleoni, J. McMinis, D.M. Ceperley, V. Olevano, and L. Delle Site,  Phys. Rev. Lett. {\bf 107}, 110402 (2011).
%\bibitem{momk2D} M. Holzmann, B. Bernu, V. Olevano, R. M. Martin, and D. M. Ceperley, Phys. Rev. B {\bf 79}, 041308 (2009).
%\bibitem{Na} S. Huotari, J. A. Soininen, T. Pylkk{\"a}nen, K. H{\"a}m{\"a}l{\"a}inen, A. Issolah, A. Titov, J. McMinis, J. Kim, K. Esler, D. M. Ceperley, M. Holzmann, and V. Olevano, Phys. Rev. Lett. {\bf 105}, 086403 (2010).
%\bibitem{TBC} C. Lin, F. H. Zong, and D. M. Ceperley, Phys. Rev. E {\bf 64}, 016702 (2001).
%\bibitem{FSE} S. Chiesa, D.M. Ceperley, R.M. Martin, and M. Holzmann, Phys. Rev. Lett. {\bf 97}, 076404 (2006).
%\bibitem{FiniteSize} M. Holzmann, R.C. Clay III, M. A. Morales, N.M. Tubman, D. M. Ceperley, and C. Pierleoni, Phys. Rev. B {\bf 94}, 035126 (2016).
%\bibitem{Pierleoni2016} C. Pierleoni, M. A. Morales, G. Rillo, M. Holzmann, and D. M. Ceperley, Proc. Natl. Acad. Sci. USA, 10.1073/pnas.1603853113 (2016).
%\bibitem{SCCS2017} C. Pierleoni, M. Holzmann, and D.M. Ceperley, SCCS 2017 Conference proceedings (2017); cond-mat/1711.00702. 
%\bibitem{Mazzola2015} G. Mazzola and S. Sorella,  Phys. Rev. Lett. {\bf 114}, 105701 (2015).

%\end{thebibliography}

\end{document}